\documentclass[11pt,a4paper,twoside,groupcitations]{article}
\usepackage[T1]{fontenc}
\usepackage[ansinew]{inputenc}
\usepackage[english]{babel}
\usepackage{amsfonts}
\usepackage{amsmath}
\usepackage{bm}
\usepackage{array}
\usepackage{amsthm}
\usepackage{amssymb}
\usepackage{graphicx}
\usepackage{subfigure}
\usepackage{braket}
\usepackage{eucal}
\usepackage{verbatim}
\usepackage[table]{xcolor}
\usepackage{caption}
\usepackage{cite}
\usepackage{textcomp}
\usepackage{multicol}
\usepackage{tikz}
\usetikzlibrary{positioning,arrows}
\usetikzlibrary{decorations.pathmorphing}
\usetikzlibrary{decorations.markings}
\usetikzlibrary{calc,decorations.markings}
\usetikzlibrary{arrows,shapes}
\usetikzlibrary{matrix,arrows}
\usepackage{pgfplots}
\usepackage{xparse}
\definecolor{jade}{HTML}{00A86B}
\newcommand{\be}{\begin{eqnarray}}
\newcommand{\ee}{\end{eqnarray}}
\renewcommand{\d}{\mbox{${\rm d}$}} %d differenziale non corsivo in math mode

\newcommand{\gn}{G_{\rm N}}
 \newcommand{\ga}{G_{(1)}}

\newcommand{\brh}{\bar{r}_{\rm H}}

\newcommand{\bt}{\bar{t}}
\newcommand{\bx}{\bar{x}}

\newcommand{\br}{\bar{r}}
\newcommand{\bA}{\bar{A}}
\newcommand{\bB}{\bar{B}}
\newcommand{\bV}{\bar{V}}
\newcommand{\bxh}{\bar{x}_{\rm H}}
\DeclareMathOperator\arctanh{arctanh}
\DeclareMathOperator\sech{sech}
\title{\bf Newtonian approximation in $(1+1)$ dimensions}
\author{Roberto~Casadio$^{ab}$\thanks{E-mail: casadio@bo.infn.it},
$\ $
Octavian~Micu$^c$\thanks{E-mail: octavian.micu@spacescience.ro}
$\ $
and
Jonas~Mureika$^{de}$\thanks{E-mail: jmureika@lmu.edu}
\\
\\
$^a${\it Dipartimento di Fisica e Astronomia, Universit\`a di Bologna}
\\
{\it via Irnerio~46, 40126 Bologna, Italy}
\\
\\
$^b${\it I.N.F.N., Sezione di Bologna, I.S.~FLAG}
\\
{\it viale B.~Pichat~6/2, 40127 Bologna, Italy}
\\
\\
{\it $^b$Institute of Space Science, Bucharest, Romania}
\\
{\it P.O. Box MG-23, RO-077125 Bucharest-Magurele, Romania}
\\
\\
{\it $^d$Department of Physics, Loyola Marymount University}
\\
{\it Los Angeles, California, USA}
\\
\\
{\it $^e$Kavli Institute for Theoretical Physics}
\\
{\it University of California Santa Barbara}
\\
{\it Santa Barbara, California, USA}
}
\begin{document}
\maketitle
\begin{abstract}
We study the possible existence of a Newtonian regime of gravity in $1+1$ dimensions, considering metrics
in both the Kerr-Schild and conformal forms.
In the former case, the metric gives the exact solution of the Poisson equation in flat space, but the weak-field
limit of the solutions and the non-relativistic regime of geodesic motion are not trivial.
We show that using harmonic coordinates, the metric is conformally flat and a weak-field expansion is straightforward.
An analysis of the non-relativistic regime of geodesic motion remains non-trivial and the weak-field potential only
satisfies the flat space Poisson equation approximately. 
\end{abstract}
\section{Introduction and motivation}
\setcounter{equation}{0}
\label{S:intro}
It has been almost three decades since 't Hooft first proposed that dimension reduction -- a decrease in the number
of spatial dimensions for ultra-high energy systems -- should be an expected feature of quantum gravity~\cite{thooft}.
Since then, high-energy dimensional reduction has been found to be a natural property
of disparate approaches to quantum gravity, including loop quantum gravity~\cite{lqg}, 
causal dynamical triangulations (CDT)~\cite{cdts}, asymptotically safe
gravity~\cite{LaR05,ReS11}, noncommutative geometry~\cite{MoN10},
vanishing dimensions \cite{Stojkovic:2013xcj},
multi-fractal geometry~\cite{Cal12}, and modified dispersion relations~\cite{AAG13+}.
\par
The nature of this dimensional reduction and its underlying mechanism are still unknown,
though there are several interpretations offered in the literature.
In the CDT framework, the effective dimension is a statistical one that results from a random walk
analysis~\cite{cdts}.
In the vanishing (or evolving) dimensions scenario, high-energy systems become progressively
more sensitive to an interlaced ``scaffolding'' of lower dimensions structures,
which can have consequences for both high energy physics and early universe cosmology~\cite{Mureika:2011bv}. 
A model~\cite{cmn,cmmn} inspired by the Generalised Uncertainty Principle invokes a duality in the expression
for the black hole mass that allows gravitation to act as if space-time were two-dimensional in the sub-Planckian
regime, even though there is no physical change in the manifold structure.
\par
Another motivation for the present analysis is given by bootstrapped Newtonian gravity (see
Refs.~\cite{Casadio:2017cdv,Casadio:2016zpl,BootN,Giusti:2019wdx,Casadio:2019cux,Casadio:2019pli,Casadio:2019tfz,
Casadio:2020ueb,Casadio:2020mch,Casadio:2020kbc,Casadio:2021gdf,DAddio:2021xsu,Casadio:2020djs})
which is built by adding non-linear terms to the Poisson equation for the Newtonian potential in $1+3$ dimensions.
In Ref.~\cite{Casadio:2021gdf}, the exact bootstrapped Newtonian potential in the vacuum in $1+3$ dimensions~\cite{BootN}
was used to reconstruct a full space-time metric in harmonic coordinates, as this is the reference frame in which
the Newtonian regime is recovered, and orbits were then studied in Ref.~\cite{DAddio:2021xsu}.
Lower-dimensional cases were considered in Ref.~\cite{Casadio:2020djs}, where the vacuum bootstrapped Newtonian
potential was lifted to a full space-time metric in $1+1$ and $1+2$ dimensions by instead assuming it to be of the
Kerr-Schild form~\cite{Jacobson:2007tj}.
\par
Clearly, for all of the above reasons, it is important to identify in which reference frame and under which conditions
gravity in $1+1$ dimensions
admits a Newtonian approximation.
In Section~\ref{S:1+3}, we start by recalling the various steps involved in the definition of this regime in General Relativity
in $1+3$ dimensions, namely the conditions under which 1) the Einstein field equations for a static source reduce to the flat
space Poisson equation for the Newtonian potential and 2) the geodesic motion of massive particles
reproduces Newton's law.
We shall then analyse gravity in $1+1$ dimensions in Section~\ref{S:1+1}, where we will employ both the
Kerr-Schild-like coordinates used in Ref.~\cite{Casadio:2020djs} and harmonic coordinates, in which
the metric is in conformally flat form.
Concluding remarks will then be presented in Section~\ref{S:conc}.
\section{Linearised Einstein theory and Newtonian potential in $1+3$ dimensions}
\setcounter{equation}{0}
\label{S:1+3}
For metric perturbations around flat space-time,
\be
g_{\mu\nu}
=
\eta_{\mu\nu}
+
\epsilon\,h_{\mu\nu}
\ ,
\ee
we can expand all terms in the Einstein tensor up to order $\epsilon$ and,
after some algebra, the linearised Einstein field equations read~\footnote{We use units with $c=1$.}
\be
L^{\alpha\beta}_{\ \ \mu\nu}\,
h_{\alpha\beta}
&\!\!\equiv\!\!&
-\Box h_{\mu\nu}
+\eta_{\mu\nu}\,\Box h 
+\partial_\mu\partial^\lambda h_{\lambda\nu}
+\partial_\nu\partial^\lambda h_{\lambda\mu}
-\eta_{\mu\nu}\,\partial^\lambda\partial^\rho h_{\lambda\rho}
-\partial_\mu\partial_\nu h
\nonumber
\\
&\!\!=\!\!&
16\,\pi\, \gn \, T_{\mu\nu}
\ ,
\quad
\label{EinstField}
\ee
in which $L^{\alpha\beta}_{\ \ \mu\nu}$ is the Lichnerowicz operator and
we consistently expanded the source energy-momentum tensor as
\be
T_{\mu\nu}^{\rm M}
=
T^{(0)}_{\mu\nu}
+
\epsilon\,T_{\mu\nu}
\ee
and used the fact that in Minkowski $T^{(0)}_{\mu\nu}=0$.
\par
We can now take advantage of the invariance of the Einstein equations under diffeomorphisms,
\be
h_{\mu\nu}
\to
\bar h_{\mu\nu}
=
h_{\mu\nu}
-
\left(\xi_{\mu,\nu}
+\xi_{\nu,\mu}\right)
\ ,
\ee
and the fact that $T_{\mu\nu}$ and $h_{\mu\nu}$ only contain six independent degrees of freedom
to impose four conditions in order to simplify the above Eq.~\eqref{EinstField}.
A famous example is given by the harmonic or de~Donder gauge condition which,
on the flat space-time, can be written as
\be
2\,\partial^\mu \bar h_{\mu\nu}
-
\partial_\nu \bar h
=
0
\ .
\label{ddgg}
\ee
In particular, the de~Donder gauge condition can likewise be used to determine
a vector field $\xi^\mu$ such that the perturbation $\bar h_{\mu\nu}$ satisfies~\eqref{ddgg}, or
\be
2\,\Box\xi_\nu
=
2\,\partial^\mu h_{\mu\nu}
-
\partial_\nu h
\ .
\label{eq:xi}
\ee
Perturbations satisfying the de~Donder gauge are usually referred to as transverse and traceless,
since one can introduce~\footnote{Note that the factor of $1/2$ (rather than $1/3$) in the definition~\eqref{hTT}
is reminiscent of the fact that gravitational waves are orthogonal to the light-like direction along which they propagate.}  
\be
h_{\mu\nu}^{\rm TT}
=
h_{\mu\nu}
-\frac{1}{2}\,\eta_{\mu\nu}\, h
\label{hTT}
\ee
and then Eq.~\eqref{ddgg} becomes the transversality condition,
\be
\partial^\mu
\bar h_{\mu\nu}^{\rm TT}
=
0
\ ,
\ee
in full analogy with the (transverse) Lorenz gauge.
\par
In the harmonic gauge~\eqref{ddgg}, taking the trace of the field equations~\eqref{EinstField} yields
\be
\Box \bar h
=
16\,\pi\,\gn\,T
\ ,
\label{eq:th}
\ee
where $T=\eta^{\mu\nu}\,T_{\mu\nu}$, and Eq.~\eqref{EinstField} reduces to
\be
-\Box \bar h_{\mu\nu}
+
\frac{1}{2}\,\eta_{\mu\nu}\,\Box\bar h
=
16\,\pi\, \gn\,T_{\mu\nu}
\ .
\ee
From Eq.~\eqref{eq:th}, we finally obtain the inhomogeneous equation
\be
-\Box \bar h_{\mu\nu}
=
16\,\pi\, \gn 
\left(T_{\mu\nu}-\frac{1}{2}\,\eta_{\mu\nu}\,T\right)
\ .
\label{deDonderField}
\ee
\par
In order to recover the Newtonian approximation, in addition to the above weak field limit,
we must assume that all matter in the system moves with a characteristic velocity much slower
than the speed of light in the (implicitly) chosen reference frame $x^\mu$.
In particular, we assume that the main source is static and the stress-energy tensor is accordingly
determined solely by the energy density,~\footnote{We
recall that the pressure is suppressed by a factor of $c^{-2}$ with respect to the energy density.}
\be
T^{\mu\nu}
\simeq
\delta^{\mu}_0\,\delta^\nu_0\,\rho(x^i)
=
T_{00}
=
-T
\ ,
\ee
The only relevant component of the metric is correspondingly $h_{00}$, and its time
derivatives are also assumed to be negligible, that is $h_{00}\simeq h_{00}(x^i)$.
In fact, for a static perturbation, the gauge condition~\eqref{ddgg} is always satisfied,
and the Ricci scalar reduces to 
\be
R
\simeq
-\nabla^2 h_{00}
\ .
\label{Rh00}
\ee
In this approximation, Eq.~\eqref{deDonderField} takes the very simple form
\be
\nabla^2 h_{00}
=
-8\,\pi\, \gn\, T_{00}
=
-8\,\pi\, \gn\,\rho
\ .
\label{hPoisson}
\ee
Since the Newtonian potential $V_{\rm N}$ is generated by the mass density $\rho$
according to the Poisson Equation
\be
\nabla^2 V_{\rm N}
=
4\,\pi\, \gn\, \rho
\ ,
\label{PoissonVN}
\ee
we can finally identify $h_{00}=-2\,V_{\rm N}$.
\subsection{Harmonic coordinates}
The non-perturbative form of the de~Donder gauge~\eqref{ddgg} is given by
\be
\Box x^\mu
=
0
\ .
\label{e:h}
\ee
In practice, one usually writes the metric for a static and spherically symmetric space-time as
\be
\d s^2
=
-\bar B\,\d \bt^2+\bar A\,d \br^2+\bar r^2\,\d\Omega^2
\ .
\label{g3}
\ee
where $\bar A=\bar A(\bar r)$, $\bar B=\bar B(\bar r)$ and $\bar r$ is the areal radius.
One then has $x^0=t$ and~\cite{weinberg} 
\be
x^1
&\!\!=\!\!&
r(\bar r)\,\sin\theta\,\cos\phi
\nonumber
\\
x^2
&\!\!=\!\!&
r(\bar r)\,\sin\theta\,\sin\phi
\\
x^3
&\!\!=\!\!&
r(\bar r)\,\cos\theta
\ ,
\nonumber
\ee
where the (invertible) function $r=r(\bar r)$ must be determined as a solution of Eq.~\eqref{e:h},
which explicitly becomes the one condition
\be
\frac{\d}{\d\bar r}
\left(
\bar r^2\,\sqrt{\frac{\bar B}{\bar A}}\,\frac{\d r}{\d \bar r}
\right)
=
2\,\sqrt{\bar A\,\bar B}\,r
\ .
\ee
We will then refer to $r$ as the ``harmonic'' radius for simplicity.~\footnote{It is important to remark that the
above equation only makes sense where $\bar A\,\bar B>0$.}
\par
The next assumption required by the weak field approximation is that
\be
\bar B(\bar r(r))
=
1+2\,V(r)
=
B(r)
\ ,
\ee
where $V$ is now the gravitational potential of choice and the last equality follows from $x^0=t$
being harmonic in a static space-time.
For example, the vacuum Schwarzschild metric is given by
\be
\bar B(\br)
=
1-\frac{2\,\gn\,M}{\bar r}
=
1-\frac{2\,\gn\,M}{r+\gn\,M}
=
B(r)
\ ,
\ee
with
\be
\br
=
r+\gn\,M
=
r
+
\frac{\brh}{2}
\ .
\ee
The weak-field regime is then given by $\br\sim r\gg 2\,\gn\,M$ in which the (radial) geodesic
equation for massive objects~\cite{weinberg,Casadio:2019oqc},~\footnote{Dots will always denote derivatives with
respect to the proper time and $J$ is the angular momentum.} 
\be
\dot \br^2
-\frac{2\,\gn\,M}{\br}
+
\frac{J^2}{\br^2}
-\frac{\gn\,M\,J^2}{\br^3}
=
E^2
-1
\ ,
\ee
reduces to the Newtonian conservation equation
\be
\frac{1}{2}\,u^2
+
V_{\rm N}
+
\frac{J^2}{2\,r^2}
\simeq
E_{\rm N}
\ ,
\ee
with $u=\d r/\d t=\dot r/\dot t$ and
\be
V_{\rm N}
\simeq
\frac{B-1}{2}
\simeq
-\frac{\gn\,M}{r}
\ll
1
\ .
\ee
\par
One interesting property of the Schwarzschild metric is that it is of the Kerr-Schild form,
that is~\cite{Jacobson:2007tj}
\be
\bar A\,\bar B=1
\ .
\label{e:ks}
\ee
This extra condition ensures that $\bA$ also changes sign at the zero $\br=\brh$ of $\bB$,
so that $\br=\brh$ is a horizon.  
\subsection{Horizon and thermodynamics} 
From $\bar r_{\rm H}$ we can then compute thermodynamic quantities.
In particular, the black hole temperature is given by
\be
T_{\rm H}
=
\frac{\hbar\,\kappa}{2\,\pi}
\ ,
\label{THgen}
\ee
where $\kappa$ is the horizon surface gravity, which can be computed from the time-like Killing
vector $k^\mu=(1,0,0,0)$ as~\cite{wald}~\footnote{The same expression is found in different dimensions.}
\be
\kappa^2
=
-\frac{1}{2}
\left(\nabla^\mu k^\nu\right)
\left(\nabla_\mu k_\nu\right)
=
\frac{\bar B'\,^2}{4\,\bar A\,\bar B}
\ .
\label{kappa2}
\ee
We recall that Hawking's result~\cite{hawking} is given by
$T_{\rm H}=\hbar/8\,\pi\,\gn\,M\simeq 4\,\hbar/100\,\gn\,M$.
\section{Gravity in $1+1$ dimensions}
\label{S:1+1}
\setcounter{equation}{0}
In addition to matter fields, all metrics in two-dimensional space-time require the presence of a dilaton field $\psi$
to assist the dynamics.
This model (dubbed $R=T$) was originally considered and studied extensively in \cite{mann1,mann2,mann3,mann4,mann5,mann6,mann7,mann8,mann9} (and additional references therein). The action is
\be
S_2
=
\int 
d^2x \,\sqrt{-g}
\left\{\frac{1}{16\,\pi\,G_{(1)}}
\left[\psi\, R
+\Lambda
+\frac{1}{2}(\nabla \psi)^2
\right]
+{\cal L}_m
\right\}
\ ,
\label{s2dilaton}
\ee
where $\psi$ is the aforementioned dilaton.
Varying (\ref{s2dilaton}) with respect to $\psi$ and $g_{\mu\nu}$ gives
the respective field equations
\be
R
=
\Box\psi
\label{vareq1}
\ee
and 
\be
\frac{1}{2}\,\nabla_\mu \psi \,\nabla_\nu \psi
- g_{\mu \nu} \left(\frac{1}{4}\,\nabla^\lambda \psi \,\nabla _\lambda \psi - \Box \psi\right)
-\nabla_\mu \nabla_\nu \psi
%&&
%\nonumber
%\\
=
8\,\pi\, G_{(1)}\, T_{\mu \nu}
+ \frac{\Lambda}{2}\, g_{\mu \nu}
\ ,
\qquad
&&
\label{vareq2}
\ee
where the stress-energy tensor is defined by
\be
T^{\mu\nu}
=
-\frac{2}{\sqrt{-g}}\,\frac{\delta {\cal L}_m}{\delta g_{\mu\nu}}
\ .
\ee
By taking the trace of the second field equation (\ref{vareq2}), one finds the left-hand side recovers (\ref{vareq1}),
which removes the dilation from the field equations.
This reduces (\ref{vareq2}) to the standard Liouville gravity form
\begin{equation}
\Box\psi
=
R
=
\Lambda
+
8\,\pi\, G_{(1)}\, T
\ .
\label{eq2}
\end{equation}
Note that the exact form of $\psi$ is irrelevant to the solution.
However, the exact Eq.~\eqref{vareq1} is already of the form~\eqref{Rh00} for a static configuration and one
thus expects the Newtonian limit is more naturally recovered in $1+1$ dimensions.
One must only assume a proper source and determine the corresponding reference frame, as we are now
going to show explicitly.
\subsection{Kerr-Schild like coordinates}
For general coordinates in which the metric is static and diagonal, 
\be
\d s^2
=
-\bB\,\d \bt^2+\bA\,d \bx^2
\ ,
\label{gAB}
\ee
the Ricci scalar reads
\be
R
=
\frac{1}{\bA\,\bB}
\left[
\frac{(\bB')^2}{2\,\bB}
+
\frac{\bA'\,\bB'}{2\,\bA}
-\bB''
\right]
\ ,
\ee
where a prime denotes differentiation with respect to the yet unspecified coordinate $\bx$. 
If we next assume that the metric is also of the Kerr-Schild form~\eqref{e:ks}, that is $\bA=\bB^{-1}$,
we then find 
\be
R=-\bB''
\ .
\ee
For a perfect fluid with energy-momentum tensor $T^\mu_{\ \nu}={\rm diag}\left[-\rho,p\right]$,
Eq.~\eqref{eq2} becomes
\be
\frac{\d^2\,\bB}{\d\,\bx^2} + \Lambda
=
8\, \pi\, G_{(1)}
\left( \rho - p\right)
\ .
\ee
Let us then consider a single point particle located at $\bx=0$, for which we can neglect the pressure $p$
and the energy density reads
\be
\rho
=
\frac{M}{2\,\pi}\,\delta(\bx)
\ .
\ee
This results in the flat space Poisson equation
\be
\frac{\d^2\,\bB}{\d\,\bx^2} + \Lambda  = 4\, \ga\,M \delta(\bx)
\ .
\label{Eeqa}
\ee 
Assuming $\bB$ is continuous and symmetric around $\bx=0$, the solution is found to be
\be
\bB
=
2\,\ga\,M\,|\bx|
-\frac{\bx^2}{\ell^2}
-C
\ ,
\label{11metric_gen}
\ee
where $C$ is an integration constant and we defined $\Lambda= \ell^{-2}$.
\par
For $C>0$ and $\ell^2\,G_{(1)}^2\,M^2>C$, the geometry contains both a black hole horizon with size
\be
\bxh
=
\ell
\left(
\ell\,G_{(1)}\,M
-\sqrt{\ell^2\,G_{(1)}^2\,M^2-C}
\right)
\label{rh1-1}
\ee
and a cosmological horizon with size
\be
\bx_\Lambda
=
\ell
\left(
\ell\,G_{(1)}\,M
+\sqrt{\ell^2\,G_{(1)}^2\,M^2-C}
\right)
\ .
\label{rL1-1}
\ee
Instead, there are no horizons if $\ell^2\,G_{(1)}^2\,M^2<C$ and one degenerate horizon $\bxh=\bx_\Lambda$
if $\ell^2\,G_{(1)}^2\,M^2=C$.
For $C\le 0$, the black hole horizon would be at $\bxh\le 0$ and one is left with the cosmological horizon $\bx_\Lambda$
only.
It is also important to remark that Minkowski geometry in $1+1$ dimensions is not recovered trivially.
In fact, the limit $M\sim\Lambda\to 0$ reverses the role of $\bt$ and $\bx$ as time and space for $C>0$
(since $\bB$ changes sign across the extremal configuration $\bxh=\bx_\Lambda$),
whereas $\bB$ does not change sign for $C<0$. 
\par
In the limit $\ell\gg (G_{(1)}\,M)^{-1}$, if $C>0$, one obtains
\be
\bxh
\simeq
\frac{C}{2\,G_{(1)}\, M}
\ ,
\label{11horizonH}
\ee
which scales as the inverse of the mass,
and
\be
\bar r_\Lambda
\simeq
{2\,\ell^2\,G_{(1)}\, M}
=
C\,\frac{\ell^2}{\bxh}
\ .
\label{11horizonL}
\ee
According to Eq.~\eqref{THgen}, this gives a Hawking temperature that runs linearly with the mass,
\be
T_{(1+1)}
\sim
\bB'(\bxh)
\sim
G_{(1)}M
\ee
and an entropy that is consequently logarithmic,
\be
S_{(1+1)} \sim G_{(1)}^{-1}\ln\left(\frac{M}{M_*}\right)
\ee
Here, $M_*$ is some new fundamental (minimal) mass scale relevant to the model.
\par
\begin{figure}[t]
\centering
\includegraphics[width=10cm]{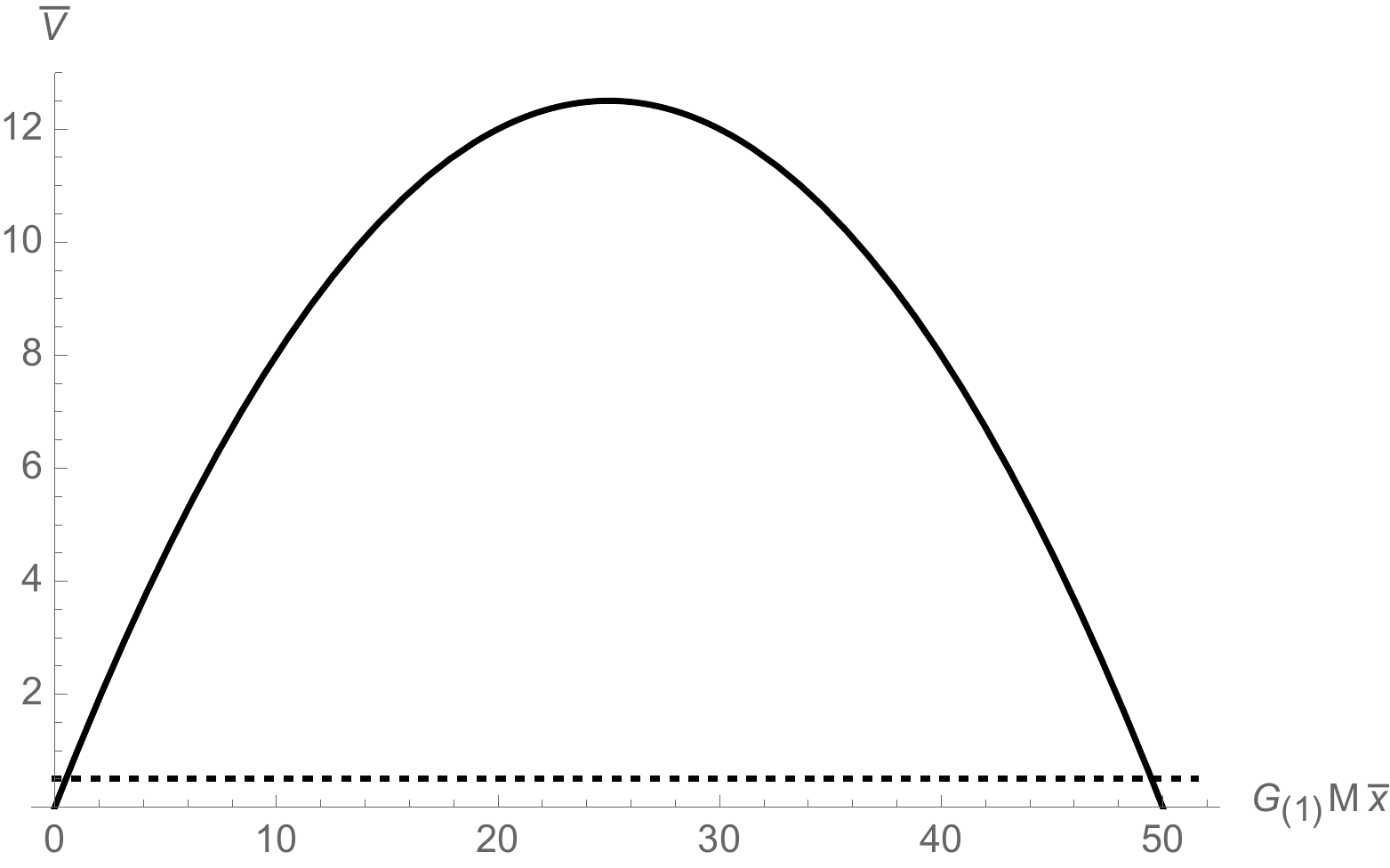}
\caption{Potential $\bV$ for $\ell=5\left(G_{(1)}\,M\right)^{-1}$ and $C=1$, 
between the two horizons $\bxh\simeq 0.5\,\left(G_{(1)}\,M\right)^{-1}$ and
$\bar x_\Lambda\simeq 50 \left(G_{(1)}\,M\right)^{-1}$, corresponding to $\bV=1/2$ 
(dotted line).}
\label{bVPlot}
\end{figure}

Geodesic motion can be easily obtained from the Lagrangian
\be
2\,L
=
\bB\,\dot\bt^2
-
\bB^{-1}\,\dot\bx^2
=
\varepsilon
\ ,
\label{geo}
\ee
where $\varepsilon=0$ for light (with the dot denoting derivatives with respect to an affine variable)
and $\varepsilon=1$ for massive particles (and the affine parameter becomes the proper time).
Since $\bB$ does not depend on $\bt$, one finds the constant of motion
\be
\bar K
=
\bB\,\dot\bt
\ .
\ee
Eq.~\eqref{geo} then reads
\be
\dot\bx^2
+
\varepsilon\,\bB
=
\bar K^2
\ .
\label{geoKS}
\ee
For $\varepsilon=1$, this equation can be written as the ``energy'' conservation equation
\be
\frac{1}{2}\,\dot\bx^2
+G_{(1)}\,M\,\bx
-
\frac{\bx^2}{2\,\ell^2}
=
\frac{C+\bar K^2}{2}
\equiv
\bar E
\ ,
\label{totE}
\ee
from which we can read out the potential (see Fig.~\ref{bVPlot} for particular values of the parameters)
\be
\bV
=
\frac{1}{2}
\left(
\bB + C
\right)
=
G_{(1)}\,M\,\bx
-
\frac{\bx^2}{2\,\ell^2}
\ .
\label{bV}
\ee
Note that we have not assumed any ``weak field'' condition, since gravity in $1+1$ is always
determined by a ``scalar potential'' according to Eq.~\eqref{eq2}.
Moreover, in these Kerr-Schild-like coordinates, the exact solution~\eqref{bV} satisfies
the Poisson equation in flat space 
\be
\frac{\d^2\,\bV}{\d\,\bx^2}
=
2\, \ga\,M \delta(\bx)
-\frac{\Lambda}{2}
\ ,
\label{EbV}
\ee
which is part of the definition of the Newtonian regime. 
\par
\begin{figure}[t]
\centering
\includegraphics[width=10cm]{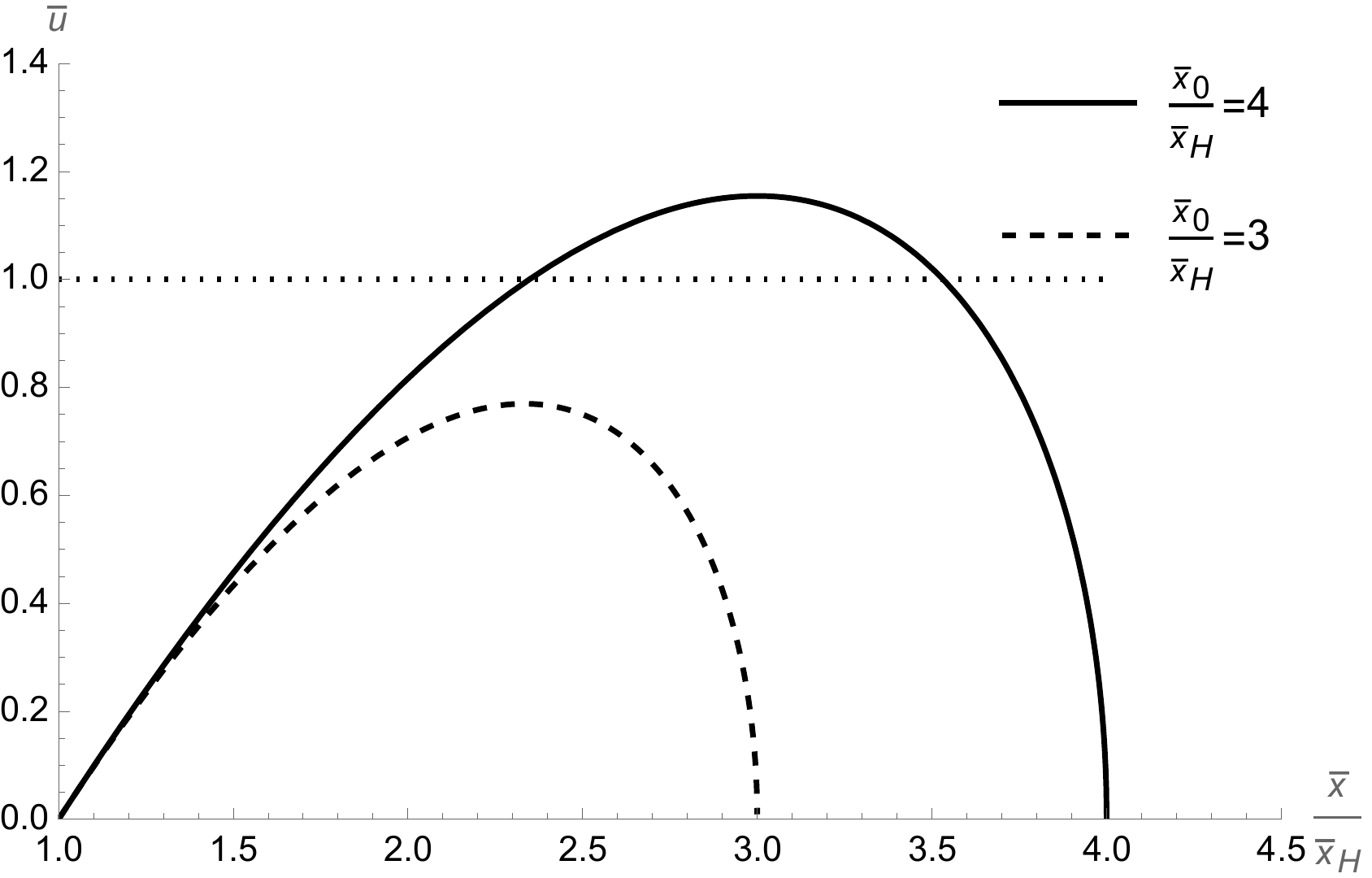}
\caption{Velocity $\bar u$ in Eq.~\eqref{uCp} for different values of $\bx_0/\bxh$.
The non-relativistic part of the trajectory occurs for $u\ll 1$ (see dotted line).}
\label{newtP}
\end{figure}

The complete Newtonian approximation also requires that the motion of massive particles be non-relativistic in the
chosen reference frame, that is $\bar u=|\dot\bx/\dot\bt|\ll 1$.
This condition simplifies considerably if we let the cosmological constant vanish, {\it i.e.}~in the limit
$\ell\rightarrow \infty$ such that $\bV=G_{(1)}\,M\,\bx$.
For $C>0$, the geometry contains the black hole horizon~\eqref{11horizonH}.
In this case, we set $C=1$ for simplicity and trade $E$ for the maximum value of $\bx$, at which the velocity $\dot\bx=0$,
\be
\bx_0
=
\frac{\bar E}{G_{(1)}\,M}
=
2\,\bxh\,\bar E
\ .
\ee
We then find that the portion of the trajectory $\bxh<\bx\le \bx_0$ is non-relativistic in the region where
\be
\bar u^2
=
\frac{\bx_0-\bx}{\bx_0-\bxh}
\left(
\frac{\bx}{\bxh}
-1
\right)^2
\ll
1
\ .
\label{uCp}
\ee
In Fig.~\ref{newtP}, two examples are plotted:
the trajectory starting with zero velocity at $\bx_0=4\,\bxh$ reaches $\bar u\simeq 1$ for 
$\bx\simeq 3.6\,\bxh$, whereas the trajectory which starts at $\bx_0=3\,\bxh$ remains
non-relativistic all the way to $\bxh$.
Note that 
\be
\bar u
\sim
\bx-\bxh
\ ,
\ee
so that $u\to 0$ for $\bx\to\bxh$ due to the usual gravitational redshift.
In $1+3$ dimensions, this portion of the trajectory occurs in the strong field regime.
However, there is no clear notion of strong field in $1+1$ because Eq.~\eqref{EbV} is exact
and the limit $M\sim\Lambda\to 0$ is not trivial for $C>0$, as we discussed above.
\par
\begin{figure}[t]
\centering
\includegraphics[width=10cm]{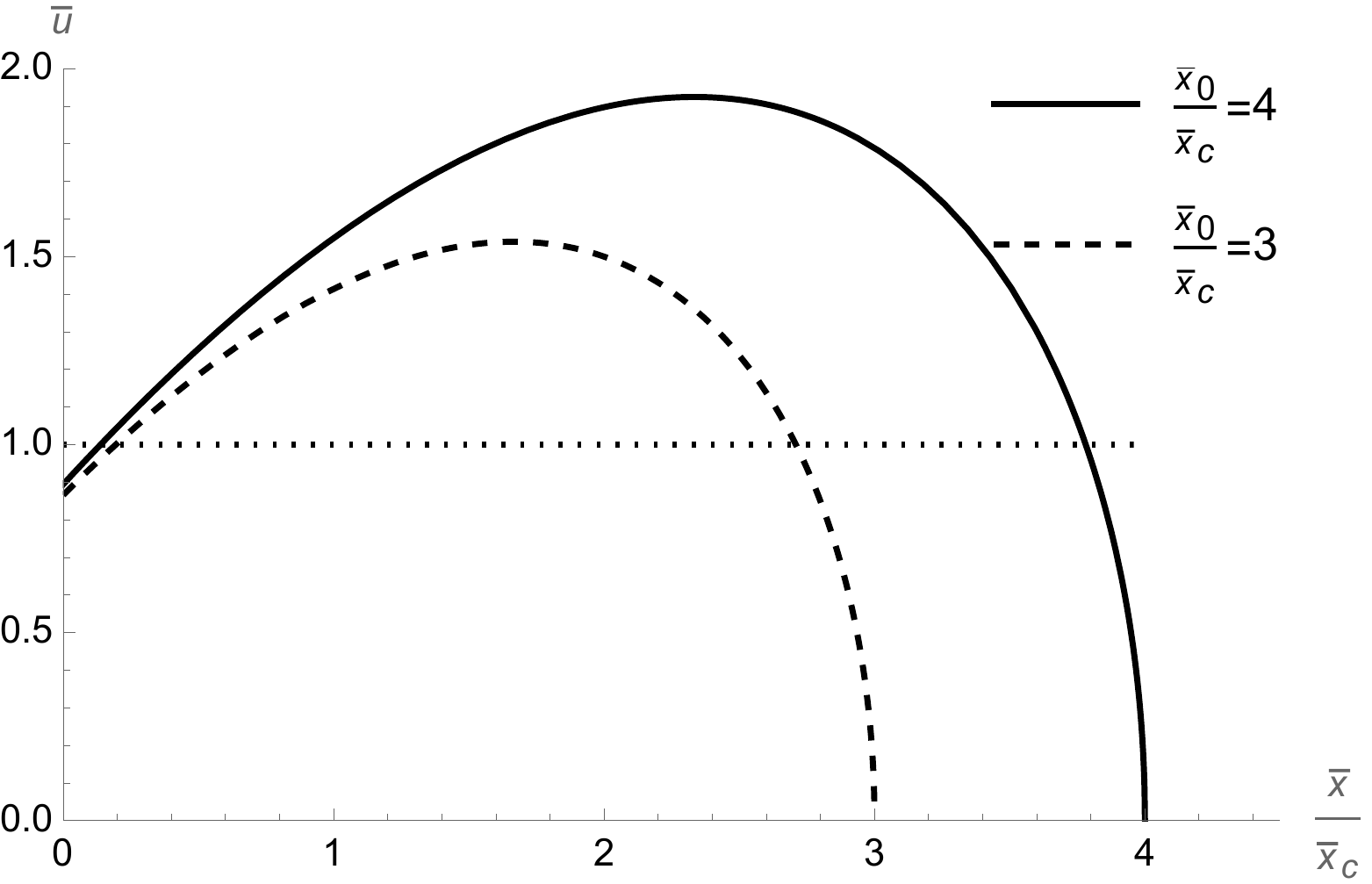}
\caption{Velocity $\bar u$ in Eq.~\eqref{uCn} for different values of $\bx_0/\bx_{\rm c}$.
The non-relativistic part of the trajectory occurs for $u\ll 1$ (see dotted line).}
\label{newtN}
\end{figure}

On the other hand, for $C<0$, there is no black hole horizon. If we set $C=-1$ for simplicity, we obtain the non-relativistic condition
\be
\bar u^2
=
\frac{\bx_0-\bx}{\bx_0+\bx_{\rm c}}
\left(
\frac{\bx}{\bx_{\rm c}}
+1
\right)^2
\ll
1
\ ,
\label{uCn}
\ee  
where $\bx_{\rm c}\equiv |\bxh|$ and $\bxh$ is given by Eq.~\eqref{11horizonH} with $C=1$. 
The above velocity is plotted in Fig.~\ref{newtN}, for the same values of $\bx_0$ and $\bxh$ used
in Fig.~\ref{newtP}.
We again see that the initial portion of the trajectory starting at $\bx_0$ is non-relativistic and the
main difference with respect to $C>0$ is that the centre $\bx=0$ is here reached with finite velocity.
\par
In order to further clarify the Newtonian regime in general, let us assume that
\be
\bB
\simeq
1+
\delta\bB
\ ,
\label{expNks}
\ee
with $|\delta\bB|\ll 1$, and rewrite the geodesic equation~\eqref{geoKS} as
\be
\frac{1}{2}\,\bar u^2
+
\bar V_{\rm N}
\simeq
\bar E_{\rm N}
\ ,
\label{NEks}
\ee
where the potential is given by
\be
\bar V_{\rm N}
\simeq
\left(\frac{1}{2}-3\,\bar E_{\rm N}\right)
\delta\bB
\ee
and
\be
\bar E_{\rm N}
\simeq
\frac{\bar K^2-1}{2\,\bar K^2}
\ .
\ee
Only for $|\bar E_{\rm N}|\ll 1$ (equivalent to $K^2\simeq 1$), one further obtains $\bar V_{\rm N}\simeq\delta\bB/2$.
This would appear to be the full Newtonian regime, in which the potential $\bar V_{\rm N}$ satisfies the flat space
Poisson equation~\eqref{Eeqa} and massive particles satisfy the non-relativistic Newtonian conservation equation~\eqref{NEks}.
\par
However, the explicit solution~\eqref{11metric_gen} does not trivially admit a weak-field expansion of the form in Eq.~\eqref{expNks},
and the above picture becomes problematic.
For $C\not=0$, the exact potential in Eq.~\eqref{NEks} is given by
\be
\bar V_{\rm N}
=
\frac{\bar V}{C^3}\,
\left[
C^2\,(6\,\bar E_{\rm N}-C^2)
+
4\,C\,(C^2-3\,\bar E_{\rm N})\,
\,\bar V
+ 
4\,(2\,\bar E_{\rm N}-C^2)\,
\bar V^2
\right]
\ ,
\ee
which again depends on the energy
\be
\bar E_{\rm N}
=
\frac{C^2\,\bar E}{2\,\bar E-C}
\ ,
\ee
and $\bar V$ was introduced in Eq.~\eqref{bV}.
Upon expanding for $|\bar V|\ll1$, we obtain
\be
\bar V_{\rm N}
\simeq
\frac{6\,\bar E_{\rm N}-C^2}{C}\,
\bar V
\ee 
and the dependence on the energy is again negligible for $|\bar E_{\rm N}|\ll C^2$,
but then one finds 
\be
\bar V_{\rm N}\simeq -C\,\bar V
\ .
\ee
For completeness, we notice that for $C=0$, one has
\be
\bar V_{\rm N}
=
2\,\bV
\left(
\frac{2}{\bar E}\,\bV
-1
\right)
\simeq
-2\,\bV
\ee
and $\bar E_{\rm N}=0$.
It thus seems that the only case which could reproduce a full Newtonian regime with an effective Newtonian
potential $\bar V_{\rm N}$ satisfying the flat space Poisson equation is given by the case $C<0$, in which we
recall that there is no black hole horizon. 
\par
This conclusion was also reached in Ref.~\cite{mann3}, where the same coordinates were employed and the
Hamiltonian dynamics of a system of particles was studied in the post-Newtonian approximation,
thus expanding on the analysis of Ref.~\cite{mann2}.
In the former paper, they conclude the coordinate condition reproduces the dynamics and geodesic equations,
derived from a canonical Hamiltonian formalism.
The authors derive a Newtonian limit from a post-linear approximation in the weak field, slow velocity limit,
and find that the equations of motion are analogous to the $(1+3)$-dimensional case.
This is in contrast to our results for $(1+1)$ dimensions.
We further note that our derivation of the Newtonian potential does not rely on any post-linear expansion,
but is a direct result of the formalism.
\par
The scenario considered herein is clearly different from the $1+3$-dimensional case, for which one knows that
harmonic coordinates must be employed in order to recover the flat space Poisson equation.
We shall next study harmonic coordinates in $1+1$ dimensions.
\subsection{Harmonic and conformal coordinates} 
The harmonic coordinate condition~\eqref{e:h} for a metric in the general form of Eq.~\eqref{gAB}
reads
\be
x'' + \frac{1}{2}\left( \frac{\bar{B}'}{\bar{B}} - \frac{\bar{A}'}{\bar{A}} \right) x'=0
\ .
\label{deq1}
\ee
If we assume that $x(\bx)=\bx$, we find that the metric is in conformally flat form.
In fact, the harmonic constraint~\eqref{deq1} simplifies to
\be
\frac{{B}^\prime}{{B}}
=
\frac{{A}^\prime}{{A}}
\ ,
\ee
which implies that $B/A$ is constant (and we omitted the bars for simplicity).
We can always rescale the coordinate $t$ in order to set this constant equal to one,
so that $A=B$ for any $B=B(x)$.
The metric finally reads 
\be
\d s^2 = B
\left(-\d t^2 + \d x^2
\right)
\ ,
\label{ds1}
\ee
which highlights the known conformally flat nature of two-dimensional space-times.
We have thus recovered the fact that harmonic coordinates are those in which this 
property is apparent.
Additionally, we also notice that the metric is not of the Kerr-Schild form in such
a reference frame (unless $B=1$ and the metric is flat).
\par
The Ricci scalar now reads
\be
R
=
-\frac{1}{B^2}
\left[
B''
-
\frac{(B')^2}{B}
\right]
\ ,
\label{RiccCF}
\ee
so that the exact field equation~\eqref{eq2} now differs from the Poisson equation in flat space.
\par
\begin{figure}[t]
\centering
\includegraphics[width=10cm]{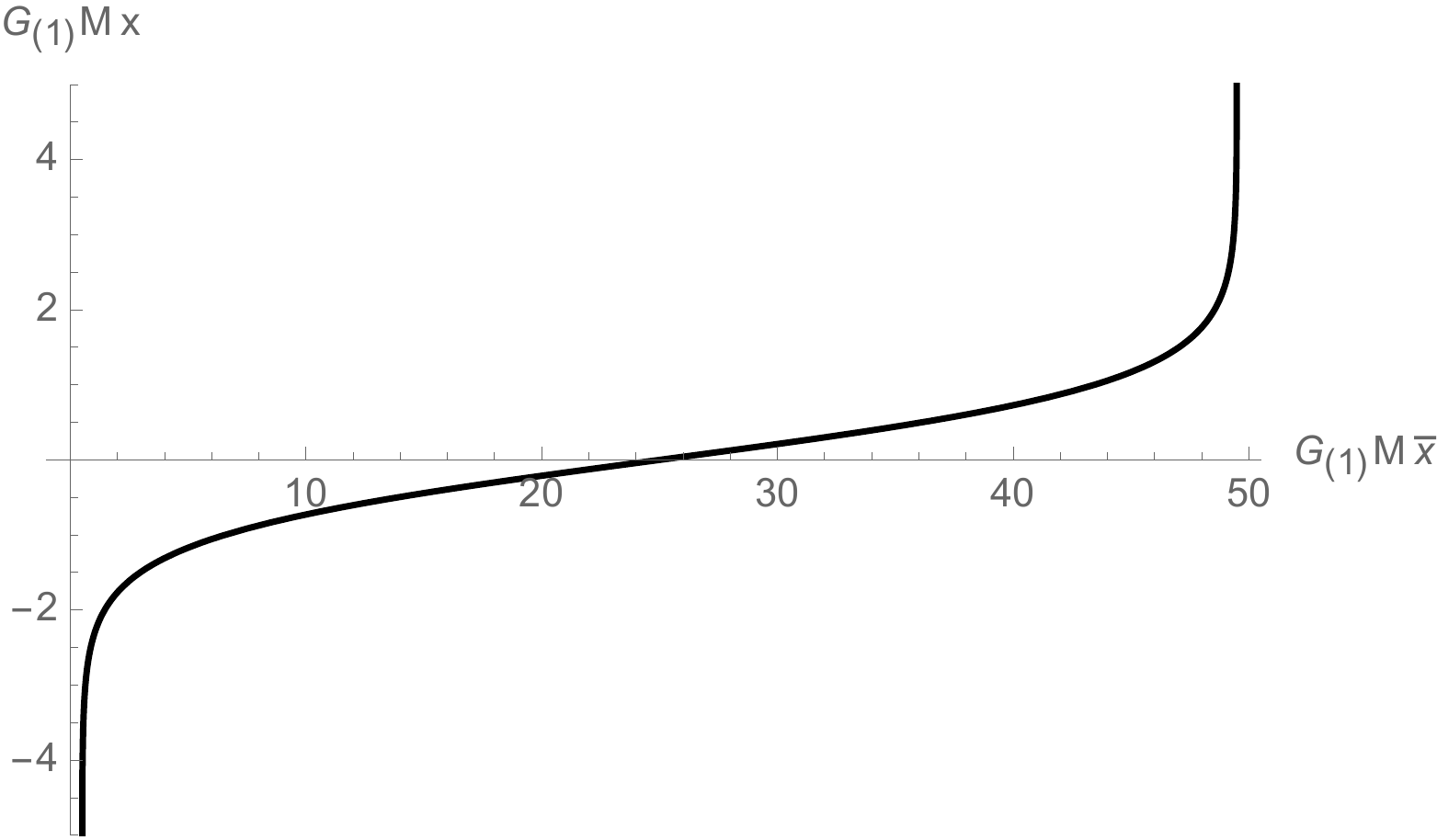}
\caption{Harmonic coordinate $x$ as a function of the Kerr-Schild-like coordinate $\bx$
for $\ell=5\left(G_{(1)}\,M\right)^{-1}$ and $C=1$.
The transformation diverges on the horizons $\bxh\simeq 0.5\,\left(G_{(1)}\,M\right)^{-1}$ and
$\bar x_\Lambda\simeq 50 \left(G_{(1)}\,M\right)^{-1}$.}
\label{xbx}
\end{figure}

When the metric is in the Kerr-Schild form in the starting coordinates $\bt=t$ and $\bx$,
the harmonic condition simplifies to
\be
x'' + \left(\frac{\bar{B}'}{\bar{B}}\right) x'=0
\ , 
\ee
which is equivalent to $x'\,\bar{B}=1$ after rescaling the coordinates by an arbitrary constant.
This equation can be solved exactly and, for $\bB$ in Eq.~\eqref{11metric_gen},
one obtains
\be
x
=
x_0
-\frac{\ell\, \arctanh \left(\frac{\ell^2\,G_{(1)}\,M-\bx}
{\ell\, \sqrt{\ell^2\,G_{(1)}^2\,M^2-C}}\right)}{\sqrt{\ell^2\,G_{(1)}^2\,M^2-C}}
\ ,
\label{xofbx}
\ee
where $x_0$ is an integration constant which we will set to zero thereafter.
The above coordinate transformation is a monotonically increasing function well-defined for $\bxh<\bx<\bar x_\Lambda$
(see Fig.~\ref{xbx}).
\par
\begin{figure}[t]
\centering
\includegraphics[width=10cm]{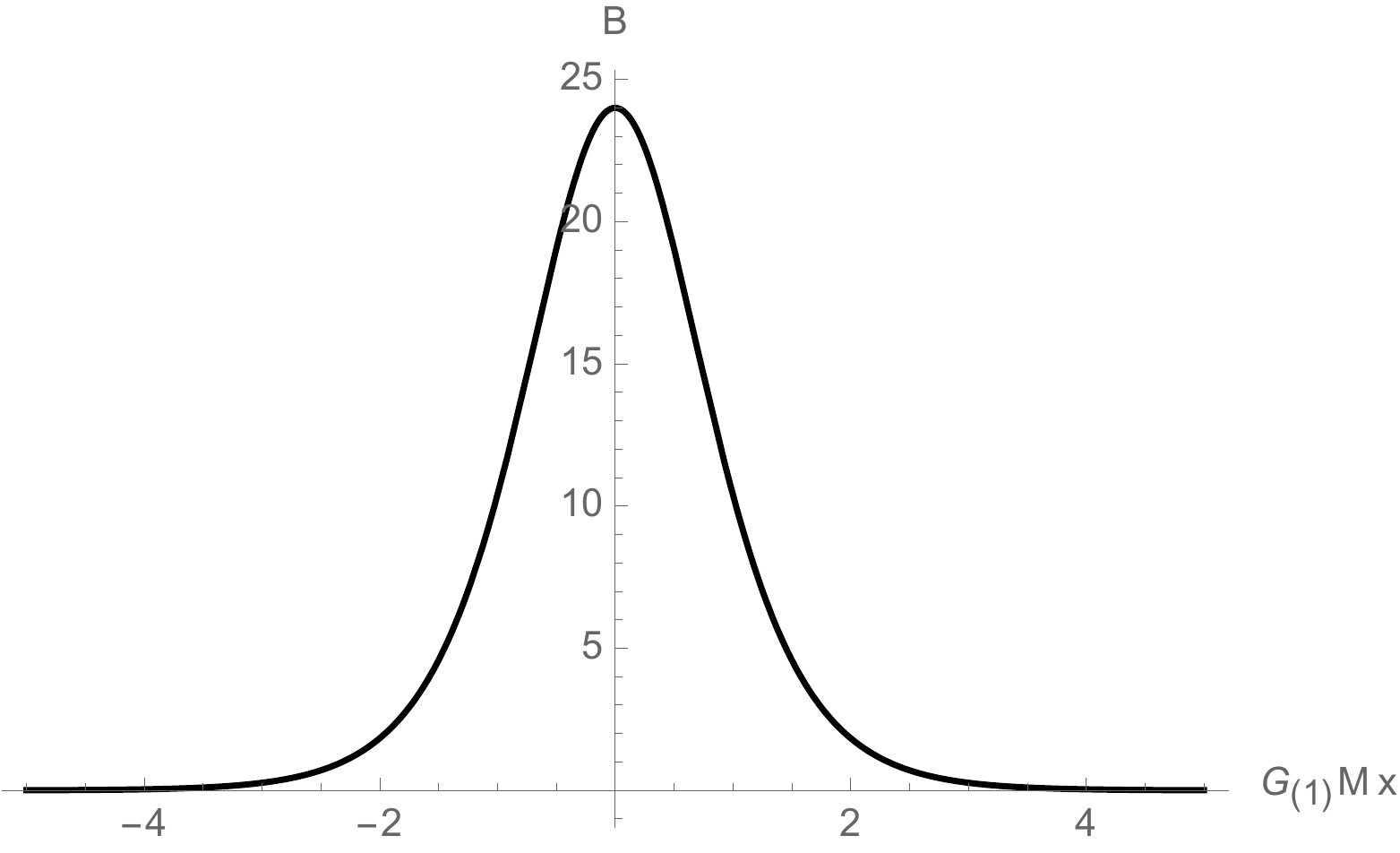}
\caption{Metric function $B$ for $\ell=5\left(G_{(1)}\,M\right)^{-1}$ and $C=1$.
Note that $\bxh$ corresponds to $x\to-\infty$ and $\bar x_\Lambda$ to $x\to+\infty$.}
\label{Bx}
\end{figure}

We can also invert the transformation~\eqref{xofbx},
\be
\bx
=
\ell\left[
\ell\,G_{(1)}\,M
+
\sqrt{\ell^2\,G_{(1)}^2\,M^2-C}\,
\tanh\left(
\sqrt{\ell^2\,G_{(1)}^2\,M^2-C}\,
\frac{x}{\ell}
\right)
\right]
\ee
and write the metric in harmonic coordinates as
\be
\d s^2 = B
\left(-\d t^2 + \d x^2
\right)
\ ,
\ee
where we recall that $\bt=t$ for static metrics and
\be
B
=
\left(\ell^2\,G_{(1)}^2\,M^2-C\right)
\left[
\sech\left(
\sqrt{\ell^2\,G_{(1)}^2\,M^2-C}\,
\frac{x}{\ell}
\right)
\right]^2
\ .
\label{Bxx}
\ee
An example is given in Fig.~\ref{Bx}, for the same parameters used in Fig.~\ref{xbx}.
\par
Geodesic motion is now described by
\be
2\,L
=
B
\left(
\dot t^2
-
\dot x^2
\right)
=
\varepsilon
\ ,
\label{geoCF}
\ee
which yields the conserved quantity
\be
K
=
B\,\dot t
\ .
\ee
For $\varepsilon=1$, we then find
\be
\dot x^2
=
\frac{K^2-B}{B^2}
\ ,
\label{geoCF}
\ee
from which it is not clear how one can read out a (Newtonian) potential.
This feature goes along with the fact the $B$ satisfies a field equation which is not of the
Poisson form in flat space.
\begin{figure}[t]
\centering
\includegraphics[width=10cm]{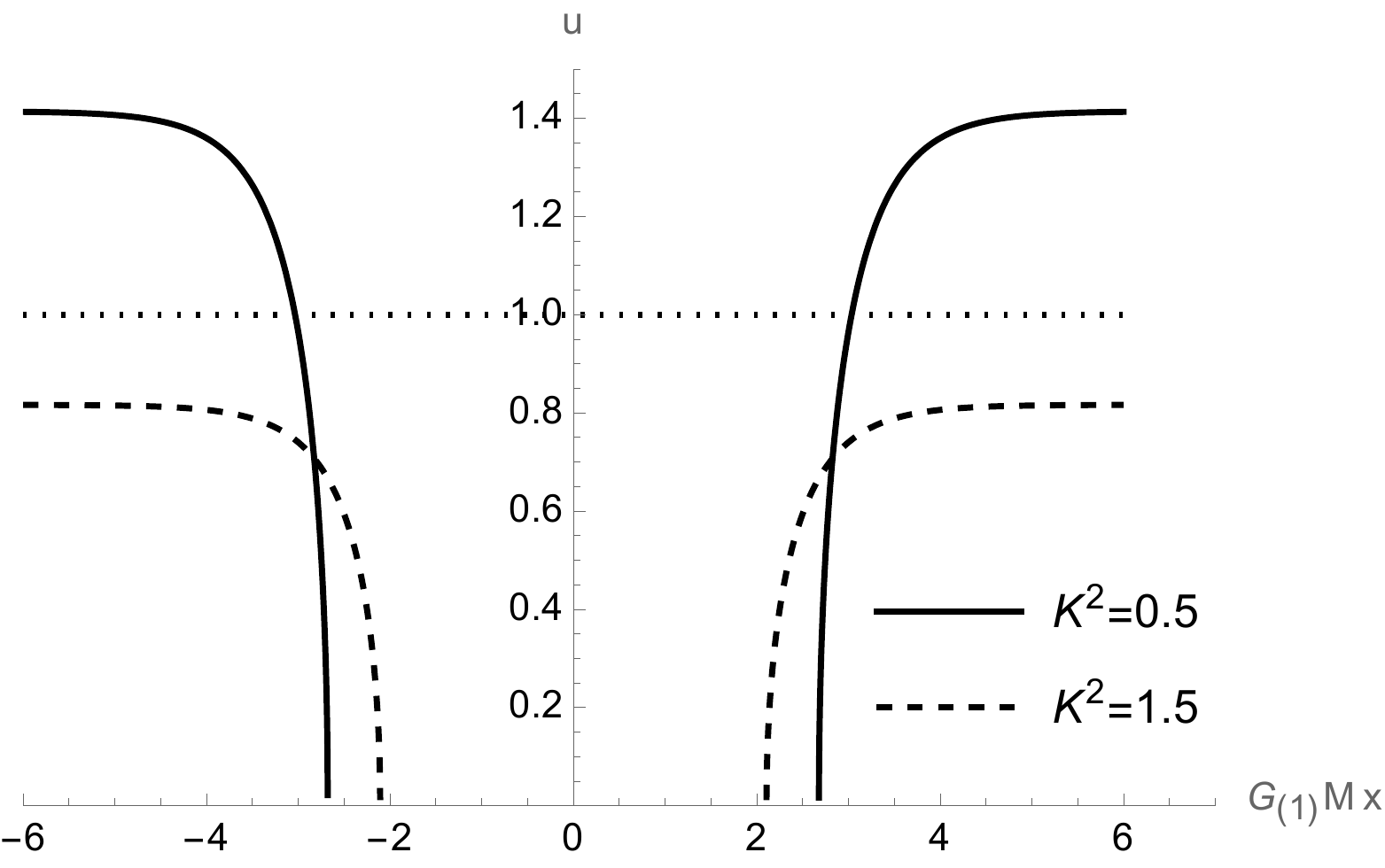}
\caption{Velocity $u$ in Eq.~\eqref{uC} for different values of $K^2$ and $B$ given in
Eq.~\eqref{Bxx} with $\ell=5\left(G_{(1)}\,M\right)^{-1}$.
The non-relativistic part of the trajectory occurs for $u\ll 1$ (see dotted line).}
\label{newtC}
\end{figure}
\par
One can still impose the condition that the motion be non-relativistic as $u\equiv |\dot x/\dot t|\ll 1$,
which reads
\be
u^2
=
\frac{\left|
K^2-B
\right|}
{K^2}
\ll
1
\ .
\label{uC}
\ee
For the potential~\eqref{Bxx} is always positive in between the two horizons (see Fig.~\ref{Bx}),
one then finds that the non-relativistic motion is restricted within a band like the one shown
in Fig.~\ref{newtC} for the same parameters used in Fig.~\ref{Bx}.
A major difference with respect to the Kerr-Schild-like coordinates, is that the velocity $u$ does not vanish 
but asymptotes to a maximum value when approaching either horizon (that is, for $|x|\to\infty$,
where $B$ vanishes).
\par
We can again investigate the Newtonian regime in general by formally assuming that
\be
B\simeq 1+\delta B
\ ,
\label{expNcf}
\ee
and rewriting the geodesic equation~\eqref{geoCF} as
\be
\frac{1}{2}\,u^2
+
V_{\rm N}
\simeq
E_{\rm N}
\ ,
\ee
where now
\be
V_{\rm N}
\simeq
\left(
1-2\,E_{\rm N}
\right)
\frac{\delta B}{2}
\ee 
and
\be
E_{\rm N}
\simeq
\frac{K^2-1}{2\,K^2}
\ .
\ee
The Newtonian limit is then equivalent to $|E_{\rm N}|\ll 1$, with $V_{\rm N}\simeq\delta B/2$.
Note that, in the approximation~\eqref{expNcf}, the Ricci scalar is approximated by
\be
R
\simeq
-2\,V_{\rm N}''
\ ,
\ee
and a fully non-relativistic Newtonian approximation is formally recovered in harmonic coordinates.
\par
By considering the explicit form of the exact solution~\eqref{Bxx}, one however finds again that it is
not trivial to identify the regime in which the approximation~\eqref{expNcf} holds.
In fact, we have already shown that the non-relativistic motion occurs when Eq.~\eqref{uC} is satisfied,
with examples given in Fig.~\ref{newtC}.
\section{Conclusions}
\setcounter{equation}{0}
\label{S:conc}
In this paper we have furthered our analysis of the $1+1$ dimensional case, expanding our understanding
of the associated Newtonian regime.
We have found that the Kerr-Schild form of the metric gives the exact solution of the Poisson equation in
flat space.
However, both the weak-field limit of the (known and exact) solutions and the non-relativistic regime of geodesic
motion are not trivial.
On the other hand, harmonic coordinates are the ones which make it explicit that the metric is conformally flat 
and a weak-field expansions looks more straightforward.
However, when considering again the exact solutions, the non-relativistic regime of geodesic motion
remain non-trivial and the the weak-field potential only satisfies the flat space Poisson equation approximately. 
\par
We also note further support for our results. Bootstrapped Newtonian gravity is a non-linear version of the
Newtonian interaction obtained by adding non-linear terms to the Poisson equation for static and spherically
symmetric sources~\cite{Casadio:2017cdv,Casadio:2016zpl,BootN,Giusti:2019wdx}. 
In Ref.~\cite{Casadio:2020djs}, assuming the Kerr-Schild form of the metric~\eqref{gAB} with $\bA=\bB^{-1}$
in $1+1$ dimensions, we reconstructed a metric $\bB=2\,\bV-1$ from the vacuum bootstrapped Newtonian potential
\be
\bV
=
\frac{1}{4\,q_V}
\left[
1-
\left(1
-
6\,q_V\,G_{(1)}\,M\,\bx
\right)^{2/3}
\right]
\ .
\label{V1out}
\ee
Upon expanding in the coupling $q_V$, one thus obtains 
\be
\bV
\simeq
G_{(1)}\,M\,\bx
+
q_V\,G_{(1)}^2\,M^2\,\bx^2
\ .
\ee
This potential hence solves the bootstrapped Newtonian equation in vacuum exactly, and further admits
a clean post-Newtonian expansion (explicitly parameterised by the coupling $q_V$).
Coupled with the results of the present paper, we have now seen that this could
indeed be a better framework to introduce the Newtonian approximation.
\par
To conclude, there does not appear to be a clear answer to a general definition of the Newtonian regime
in $1+1$ dimensions. The best one can do is to find exact solutions and analyse each one of them carefully. 
\section*{Acknowledgments}
R.C.~is partially supported by the INFN grant FLAG and his work has also been carried out in
the framework of activities of the National Group of Mathematical Physics (GNFM, INdAM).
O.M.~was supported by the Romanian Ministry of Research, Innovation, and Digitisation,
grant no.~16N/2019 within the National Nucleus Program. 
J.M.~thanks the I.N.F.N.~and the Department of Physics and Astronomy of the University of Bologna
for generous hospitality, at which some of this work was completed.
J.M.~is a KITP Scholar at the Kavli Institute for Theoretical Physics.
The KITP Scholars Program is supported in part by the National Science
Foundation under Grant No.~NSF PHY-1748958.
\end{document}